\documentclass[preprint,showpacs]{revtex4}
\usepackage{graphicx}
\include{epsf}

\begin{document}

\title{   
Genetic Toggle Switch Without Cooperative Binding
}
\author{Azi Lipshtat$^1$, Adiel Loinger$^2$, 
Nathalie Q. Balaban$^2$ and Ofer Biham$^2$}   
\affiliation{$^1$
Department of Pharmacology and Biological Chemistry,
Mount Sinai \\ School of Medicine,
New York, NY 10029, USA
\\
$^2$ Racah Institute of Physics, 
The Hebrew University, 
Jerusalem 91904, Israel 
}

\begin{abstract}

Genetic switch systems with mutual repression of two
transcription factors are studied using deterministic and 
stochastic methods.
Numerous studies have concluded 
that cooperative binding is 
a necessary condition for the emergence of bistability
in these systems.
Here we show that for a range of
biologically relevant conditions,
a suitable combination of network structure and
stochastic effects gives rise to bistability even
without cooperative binding. 

\end{abstract}


\maketitle

Recent advances in quantitative measurements 
of gene expression at the single-cell level 
\cite{Elowitz2002,Ozbudak2002}
have brought new insight on the importance
of stochastic fluctuations 
in genetic circuits
\cite{Mcadams1997}.
Populations of genetically identical cells show variability
due to fluctuations.
The role of fluctuations is enhanced due to 
the discrete nature of the transcription factors
and their binding sites, which may appear in low copy numbers
\cite{Becskei2000,Kaern2005}.
Stochastic behavior
may invoke oscillations
\cite{Vilar2002,Zhou2005}
and spatio-temporal patterns
\cite{Shnerb2000,Howard2003,Togashi2004},
which are unaccounted for by
macroscopic chemical rate equations. 
Genetic circuits with
feedback mechanisms often exhibit bistability,
namely, two distinct stable states which
can be switched either spontaneously or by an external 
signal
\cite{Atkinson2003,Ozbudak2004,Francois2005}.
To qualify as a switch, the spontaneous switching rate
must be much lower than the rates of the relevant
processes in the cell, namely transcription, translation,
binding and unbinding of transcription factors.
In particular, genetic switches such as the 
phage $\lambda$ switch, 
enable cells to adopt different fates
\cite{Ptashne1992}.
The toggle switch is
a simple genetic circuit that
consists of two proteins, $A$ and $B$,
with concentrations
$[A]$ and $[B]$, respectively, 
which negatively regulate each other's synthesis 
(by concentration we mean the average copy number
of proteins per cell).
The production of protein $A$ is 
negatively regulated by protein $B$,
through binding of $n$ copies of $B$
to the $A$ promoter (and vice versa). 
This process can be
modeled by a Hill function, which 
reduces the production rate of
$A$ 
by a factor of
$1+k[B]^n$,
where 
$k$ is a parameter and 
$n$ is the Hill coefficient
\cite{Hill}.
In case that $n=1$ the binding of a single protein is 
sufficient in order to perform the negative regulation, while
for $n>1$ the {\it cooperative binding} of two or more
proteins is required.
In numerous studies of the toggle switch system
it was concluded that cooperative binding is
a necessary condition for the emergence of 
the two distinct stable states
characteristic of a switch
\cite{Gardner2000,Cherry2000,Warren2004,Walczak2005}. 
It was also observed that in presence of cooperative binding,
stochastic effects contribute to the broadening of the parameter
range in which bistability appears
\cite{Kepler2001}.

In this letter we show 
that stochastic effects enable bistability even without cooperative binding
of the transcription factors to the operator, namely for
Hill coefficient $n=1$.
Furthermore, bistability takes place even when
the active proteins appear in high copy numbers.
These results 
emphasize the necessity of stochastic methods in the analysis of
genetic networks, even under conditions of high concentrations.

The mutual repression circuit,
referred to as the general switch
\cite{Warren2004},
is described by 
the rate equations
\begin{eqnarray}
\label{eq:A_dot} 
[\dot{A}] &=& g_A (1-[r_B]) - d_A [A]-
\alpha_0 [A] \left (1-[r_A] \right) + \alpha_1[r_A]  \nonumber \\
\label{Eq:B_dot}   
[\dot{B}] &=& g_B(1-[r_A])-d_B[B]-
\alpha_0[B]\left(1-[r_B]\right)+\alpha_1[r_B] \nonumber \\
\label{eq:rA_dot} 
[\dot{r_A}] &=& \alpha_0[A]\left(1-[r_A]\right)-\alpha_1[r_A] \nonumber \\
\label{eq:rB_dot}
[\dot{r_B}] &=& \alpha_0[B]\left(1-[r_B]\right)-\alpha_1[r_B],
\label{eq:rate}
\end{eqnarray}
\noindent
where $g_X$ 
(s$^{-1}$), 
$X=A,B$, is the maximal production rate 
of protein $X$ 
and
$d_X$ (s$^{-1}$) 
is its degradation rate.
For simplicity, we ignore the 
mRNA level and take the processes of transcription and translation as a 
single step of synthesis
\cite{mRNA}. 
The bound repressors are considered 
as separate species 
$r_X$
and their concentrations 
are given by
$[r_X]$,
providing much insight into the
repression process
\cite{Lipshtat2005}. 
Here, 
$r_A$ is a bound $A$ protein
that monitors the production of $B$, while
$r_B$ is a bound $B$ protein
that monitors the production of $A$.
Since there is a single promoter of each type, 
$0\le [r_X] \le 1$.
The parameter 
$\alpha_0$ (s$^{-1}$)
is the binding rate of proteins to the promoter
and 
$\alpha_1$ (s$^{-1}$) 
is the dissociation rate. 

It is commonly assumed that the 
binding-unbinding processes 
are much faster than other 
processes in the circuit, namely 
$\alpha_0,\alpha_1 \gg d_X,g_X$. 
This means that the relaxation times of $[r_X]$ 
are much shorter than other relaxation 
times in the circuit. 
Under this assumption,
one can take the 
time derivatives of $[r_X]$
to zero, even if the system is away from
steady state.
This brings the rate equations to the 
standard Michaelis-Menten form
\begin{eqnarray}    
\label{eq:dA}
[\dot{A}] &=& {g_A / ({1+k[B]})} - d_A [A] \nonumber \\
\label{eq:dB}
[\dot{B}] &=& {g_B / ({1+k[A]})} - d_B [B],
\end{eqnarray}    
\noindent
where 
$k=\alpha_0/\alpha_1$ 
is the repression strength.
For a given population of free $X$ repressors,
the parameter $k$  controls the value of $[r_X]$.
The limit of weak repression,
$[r_X] \ll 1$,
is obtained when
$k[X] \ll 1$,
while the limit of strong repression,
$[r_X] \simeq 1$,
is obtained for
$k[X] \gg 1$. 
These equations turn out to have one positive 
steady-state solution, thus at the level of rate equations
this system does not exhibit bistability.
For symmetric parameters, 
where 
$g_A=g_B=g$ 
and 
$d_A=d_B=d$,
this solution is
$[A] = [B] = [ (1+4kg/d)^{1/2} -1 ]/2k$.

In order to 
account for stochastic effects,  
the master equation approach
\cite{Mcadams1997,Kepler2001,Paulsson2000}
is applied.
In the master equation, the dynamic variables are the 
probabilities 
$P(N_A,N_B,r_A,r_B)$
for a cell to 
include
$N_X$ copies of free protein $X$
and 
$r_X$ copies of the bound $X$ repressor,
where $N_X=0,1,2,\dots$,  
and
$r_X = 0,1$.
The master equation for the mutual repression circuit
takes the form
\begin{eqnarray}
\label{eq:master}
&&\dot{P}(N_A,N_B,r_A,r_B) = 
 g_A \delta_{r_B,0} P(N_A-1,N_B,r_A,r_B) + g_B \delta_{r_A,0} P(N_A,N_B-1,r_A,r_B) 
\nonumber \\
&& + d_A (N_A+1) P(N_A+1,N_B,r_A,r_B) + d_B (N_B+1) P(N_A,N_B+1,r_A,r_B) 
\nonumber \\
&& - (g_A \delta_{r_B,0} + g_B \delta_{r_A,0}) P(N_A,N_B,r_A,r_B) - (d_A N_A + d_B N_B) P(N_A,N_B,r_A,r_B) 
\nonumber \\
&&+ \alpha_0 [ (N_A+1) \delta_{r_A,1} P(N_A+1,N_B,0,r_B) +  (N_B+1)\delta_{r_B,1} P(N_A,N_B+1,r_A,0)]
\nonumber \\
&&+ \alpha_1 [\delta_{r_A,0} P(N_A-1,N_B,1,r_B)
+ \delta_{r_B,0} P(N_A,N_B-1,r_A,1)]
\nonumber \\
&&- \alpha_0 ( N_A \delta_{r_A,0} +  N_B \delta_{r_B,0}) P(N_A,N_B,r_A,r_B)
-\alpha_1 (\delta_{r_A,1} + \delta_{r_B,1}) P(N_A,N_B,r_A,r_B),
\end{eqnarray}
\noindent
where 
$\delta_{i,j}=1$ for $i=j$ and $0$ otherwise.
The $g_X$ terms account for the production of proteins.
The $d_X$ terms account for the degradation of free proteins,
while the $\alpha_0$ ($\alpha_1$) terms describe the 
binding (unbinding) of proteins to (from) the promoter site. 
The average copy numbers 
$\langle X \rangle$, 
where $X=N_A,N_B,r_A,r_B$, 
are given by 
$\langle X \rangle =\sum X P(N_A,N_B,r_A,r_B)$
where the sum is over all integer values of $N_A$
and $N_B$ up to a suitable cutoffs and over
$r_A,r_B=0,1$.
Note that for distributions that are skewed or 
exhibit several peaks, the average does not reflect
the actual behavior in a single cell.

To analyze the role of fluctuations in 
this circuit we
have calculated the
probability distribution 
$P(N_A,N_B) = \sum_{r_A,r_B} P(N_A,N_B,r_A,r_B)$. 
We used the symmetric parameters 
$g=0.05$ (s$^{-1}$),
which correspond to average production time of $20$ seconds, 
and  
$d=0.005$ (s$^{-1}$) 
which means degradation 
time of 200 seconds,
in agreement with experimental results
\cite{Elowitz2000}. 
To examine a broad range of relevant values of $k$ we
performed two sets of simulations.
In the first set we chose
$\alpha_1=0.5$ (s$^{-1}$) 
and varied 
$\alpha_0$,
while in the second set we chose
$\alpha_0=0.5$ (s$^{-1}$) 
and varied 
$\alpha_1$.
We confirmed that the population of free
proteins depends only on the ratio, $k$.

Under conditions in which the promoter sites are
empty most of the time, namely
$r_X \ll 1$, the repression is weak and
the steady state solution
exhibits coexistence of $A$ and $B$ proteins in the cell. 
In this case the distribution 
$P(N_A,N_B)$
exhibits a single peak
[Fig.~\ref{fig:1}(a)].
In this case, the values of
$\langle N_A \rangle$
and
$\langle N_B \rangle$
obtained from the master equation
coincide with $[A]$ and $[B]$, obtained from
the rate equations.
For strong repression,
the distribution $P(N_A,N_B)$ 
exhibits a peak
in which the $A$ population is
suppressed and a peak in which the $B$ population
is suppressed, as expected for a bistable system. 
However, a third peak appears near the origin,
in which both populations of free proteins diminish
[Fig.~\ref{fig:1}(b)]. 
This peak represents a
dead-lock situation, caused by the 
fact that both $A$ and $B$ repressors can be bound
simultaneously, each
bringing to a halt the production of the other specie.
This result is in contrast to the rate
equations which exhibit a single solution,
$[A]=[B]$, for the entire range of parameters.
Below, we present three biologically sensible variants
of the circuit in which the third peak is suppressed, 
giving rise to a bistable switch. 

Consider the exclusive switch, where 
there is an overlap between the promoters 
of $A$ and $B$ and thus
no room for both to be occupied simultaneously.
Such a situation is encountered in nature,  
for example, in the 
lysis-lysogeny switch of phage $\lambda$
\cite{Ptashne1992}.
It was shown that 
in presence of cooperative binding,
the exclusive switch is more stable than
the general switch
\cite{Warren2004}.
This is because in the exclusive switch the access of the
minority specie to the promoter site is blocked by the
dominant specie.
Here we show that in the exclusive switch, 
stochastic effects give rise to bistability even
without cooperativity between the transcription factors.
To model this system recall that
$[r_A]$  
($[r_B]$)  
can be defined as the fraction of 
time in which the promoter 
is occupied by a bound $A$ ($B$) protein.
The fraction of time in which the promoter is vacant is
$1 - [r_A] - [r_B]$.
Incorporating this into Eq.~(\ref{eq:rate})
gives rise to the following modification:
in the $\alpha_0$ terms,
each appearance of 
$[r_A]$
or
$[r_B]$
should be replaced by 
$[r_A]+[r_B]$.
For symmetric parameters,
the resulting equations still exhibit
a single solution, 
in which 
$[A]=[B]$ 
and 
$[r_A]=[r_B]$. 
The Michaelis-Menten
equations for the exclusive switch are given by
Eqs.~(\ref{eq:dB}),
where in the first equation
$k$ is replaced by
$k/(1+k[A])$
and in the second equation it is replaced by
$k/(1+k[B])$.
To account for the discreteness
of the transcription factors and their fluctuations, 
the master equation
should be applied,
with the constraint that
$P(N_A,N_B,1,1)=0$.
It takes the form of 
Eq.~(\ref{eq:master}),
except that 
in the $\alpha_0$ and $\alpha_1$ terms,
each time 
$\delta_{r_A,j}$
($\delta_{r_B,j}$)
appears it should be multiplied by
$\delta_{r_B,0}$,
($\delta_{r_A,0}$).
In the exclusive switch, under conditions of
weak repression,
$P(N_A,N_B)$
exhibits a single peak
[Fig.~\ref{fig:2}(a)],
for which
$\langle N_A \rangle$
and
$\langle N_B \rangle$
coincide with
$[A]$ and $[B]$, respectively.
For strong repression,
the distribution $P(N_A,N_B)$ 
exhibits two peaks.
In one peak the $A$ population is
suppressed, while in the 
other peak the $B$ population
is suppressed, 
as expected for a bistable system 
[Fig.~\ref{fig:2}(b)].
The dead-lock situation is impossible in this system.

To examine the time dependence of the populations of free
proteins in a single cell, we have performed Monte Carlo 
simulations, based on the master equation for the exclusive
switch.
In Fig.~\ref{fig:3}
we present the copy numbers of free and bound 
$A$ and $B$
proteins
vs. time. The population size of the dominant specie is in the 
range of 20-60, while the minority specie is almost completely
suppressed. The typical switching time is around $10^{5}$ seconds.

Consider a different variant of the genetic switch, which
exhibits bound-repressor degradation (BRD).
Even a low degradation rate, $d_r$,
of the bound repressors tends to remove the
mutual suppression of both species, and gives
rise to a binary switch.
The rate equations that describe this circuit are
identical to 
Eq.~(\ref{eq:rate}),
except that 
a degradation term of the form
$-d_r[r_A]$
($-d_r[r_B]$)
is added to the equation for 
$[\dot{r_A}]$
($[\dot{r_B}]$).
For symmetric parameters,
the Michaelis-Menten form of these equations,
applicable in the limit of fast switching,
is given by 
Eq.~(\ref{eq:dB})
where 
$k=\alpha_0/(\alpha_1+d_r)$
and $d$ is replaced by an effective degradation rate
$d_{\rm eff}=d+ d_r k/(1+k[A])$ 
in the first equation
and by the analogous term in the second equation.
This equation exhibits a bifurcation
at $k_c=(d/d_r)(\sqrt{g}+\sqrt{d_r})/(\sqrt{g}-\sqrt{d_r})$, 
in which the
symmetric solution
$[A]=[B]$
becomes unstable, giving rise to two stable solutions
in which one specie is dominant and the other is suppressed
(Fig.~\ref{fig:4}, inset).
We thus find that in case that bound repressors exhibit
degradation, bistability appears even at the level of
rate equations.
The emergence of bistability can be attributed to the 
fact that the effective degradation rate for the minority
specie is larger than for the dominant specie, 
enhancing the difference between the population sizes.
The master equation for this circuit is
obtained by adding the term
$d_r [  (\delta_{r_A,0}-\delta_{r_A,1}) P(N_A,N_B,1,r_B)
      + (\delta_{r_B,0}-\delta_{r_B,1}) P(N_A,N_B,r_A,1)]$
to 
Eq.~(\ref{eq:master}).
This term represents transitions of the cell from 
$r_X=1$ ($X=A,B$) to $r_X=0$, 
{\em without}
changing the number of free proteins. 
The degradation of bound repressors gives rise to suppression
of the peak near the origin,  
leading to the emergence of bistability. 

A third variant of the genetic switch exhibits 
protein-protein interactions (PPI) such that
an A protein and a B protein may form an AB complex,
which is not active as a transcription factor.
This circuit
exhibits bistability within a range of parameters,
both for the rate equations and for the master equation.

We have calculated the switching time using the master equation,
for an initial state that includes only free A proteins.
The distribution 
$P(N_A,N_B)$ vs. time was calculated and
the function 
$f(t)=P(N_A>N_B)-P(N_A<N_B)$
was found to decay exponentially according to
$f(t)=\exp(-t/\tau)$, 
where $\tau$
is defined as the switching time.
In Fig.~\ref{fig:4} we present the switching time 
$\tau$,
obtained from
the master equation vs. $k$ for the exclusive switch
($\circ$)
and for the BRD switch
($\times$).
We also examined the dependence of $\tau$ on the copy
number, $N$, of the dominant specie. 
For the exclusive switch, we found that
when $d$ is varied, $\tau \sim N^2$, while in case
that $g$ is varied, $\tau \sim N$. This dependence
is weaker than found for the cooperative switch
\cite{Warren2004}.

The results presented in this paper 
(except for Fig.~\ref{fig:3})
were obtained by direct 
integration of the master equation rather than by Monte
Carlo methods
\cite{Gillespie1977}.
Direct integration is much more efficient and provides more accurate
results, without the need to accumulate statistics. 
Recent improvements in the methodology enable to use
direct integration for complex networks that involve large
numbers of active proteins
\cite{Lipshtat2004},
which will enable to go beyond elementary circuits into simulations
of complete networks. 

In contrast to previous knowledge that bistability requires cooperative
binding of transcription factors,
we have shown that 
bistability is possible 
without cooperative binding.
We have analyzed three 
variants of the genetic toggle switch, that exhibit bistability
without cooperative binding.
The first circuit 
is the exclusive switch, in which
the two promoter sites
cannot be occupied simultaneously.
The second circuit exhibits 
degradation of bound repressors, while in the third circuit
free $A$ and $B$ proteins may form a complex which is not
active as a transcription factor.
Rate equations predict a single stable state 
in the first circuit and bistability in the second and third
circuits.
However, the master equation predicts 
bistability in all the three circuits.
These findings are not limited to cases in which proteins exist in
low numbers, but are due to the low copy number
of the promoter itself.
The results presented here are expected to
have significant implications on the understanding of
non-genetic 
variability in cell populations,
and may shed new light on the way cells
differentiate despite uniform environmental conditions.

N.Q.B. was supported by the Center of Complexity of the Horowitz 
Foundation and the Bikura Program of the Israel Science Foundation.

\clearpage

\begin{figure}
\caption{
The probabilities
$P(N_A,N_B)$ 
for the general switch, 
under conditions of 
(a) weak repression ($k=0.005$) 
where there is one symmetric peak
and 
(b) strong repression ($k=50$)
where three peaks appear, one dominated by $A$, the
second dominated by $B$ and the third in which 
both species are mutually suppressed. 
The weights of the three peaks are about the same.
}
\label{fig:1}
\end{figure}

\begin{figure}
\caption{
The probabilities
$P(N_A,N_B)$ 
for the exclusive switch, 
under conditions of 
(a) weak repression ($k=0.005$) 
where there is one symmetric peak 
and 
(b) strong repression ($k=50$)
where bistability is observed.
}
\label{fig:2}
\end{figure}

\begin{figure}
\caption{
The populations of free and bound 
$A$ and $B$ proteins vs. time, obtained
from Monte Carlo simulations of the exclusive switch
with the parameters 
$g=0.2$, $d=0.005$, 
$\alpha_0=0.2$ 
and 
$\alpha_1=0.01$. 
The bistable behavior is clearly observed, where the population size 
of the dominant specie is between 20-60 and the other specie is nearly
diminished. Failed switching attempts are clearly seen.
}
\label{fig:3}
\end{figure}

\begin{figure}
\caption{
The switching time vs. the repression strength, $k$, for the
exclusive switch ($\circ$) and for the case in which bound
repressors exhibit degradation ($\times$). For the bistable range
(roughly $k>1$) 
the switching time increases as $k$ is increased.
The inset shows the steady state solution 
for $[A]$ and $[B]$ vs. $k$, 
obtained from the rate equations for the BRD switch.
Note that for the BRD switch, the parameter $\alpha_0$
varies, while $d_r=d$ and $\alpha_1=0.01$ are held fixed.
}
\label{fig:4}
\end{figure}


\begin{thebibliography}{10}
\bibitem{Hill}
Hill-function models are simplifications of rate-law equations.
When derived directly from rate laws, $n$ is expected to take only
integer values. However, when these models are used for fitting 
empirical data, $n$ is a fitting parameter which may take 
non-integer values.
\bibitem{mRNA}
Extended circuits that include the mRNA level have also been
studied. It was found that 
there is
a perfect agreement between the rate equation
results for the simplified and
the extended circuits,
and excellent agreement with only slight difference in peak
shapes in the master equation.
\end{thebibliography}
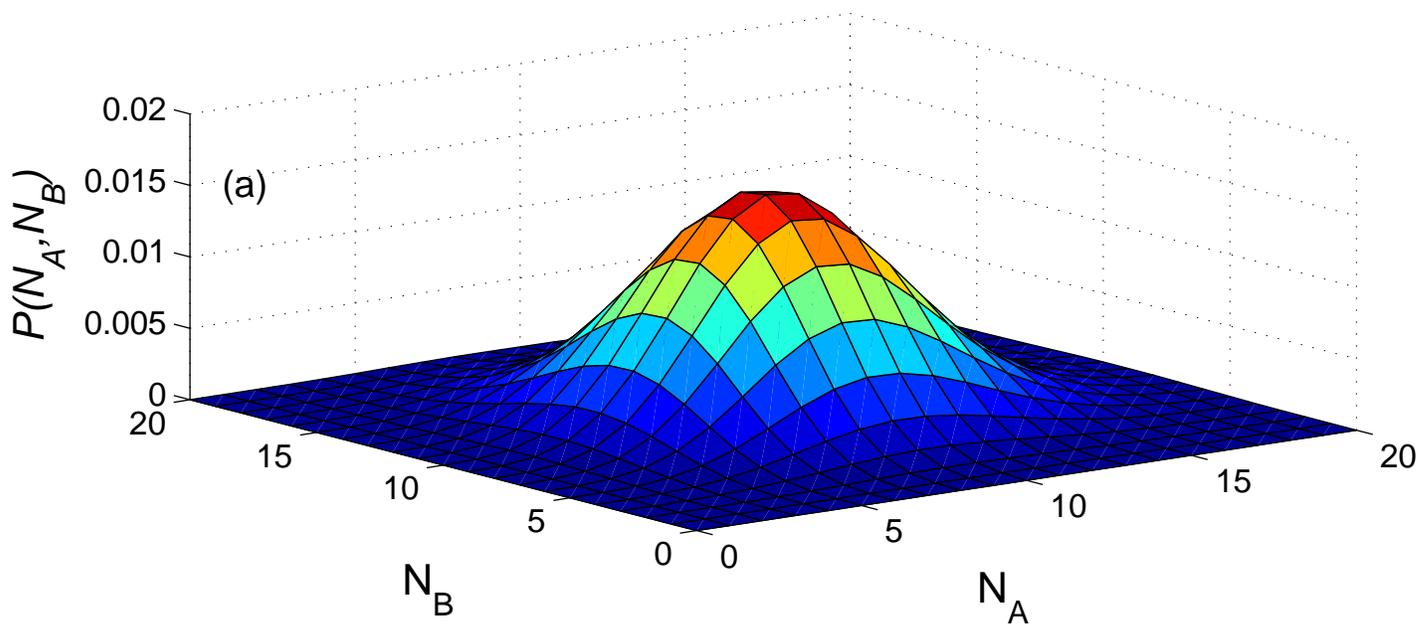
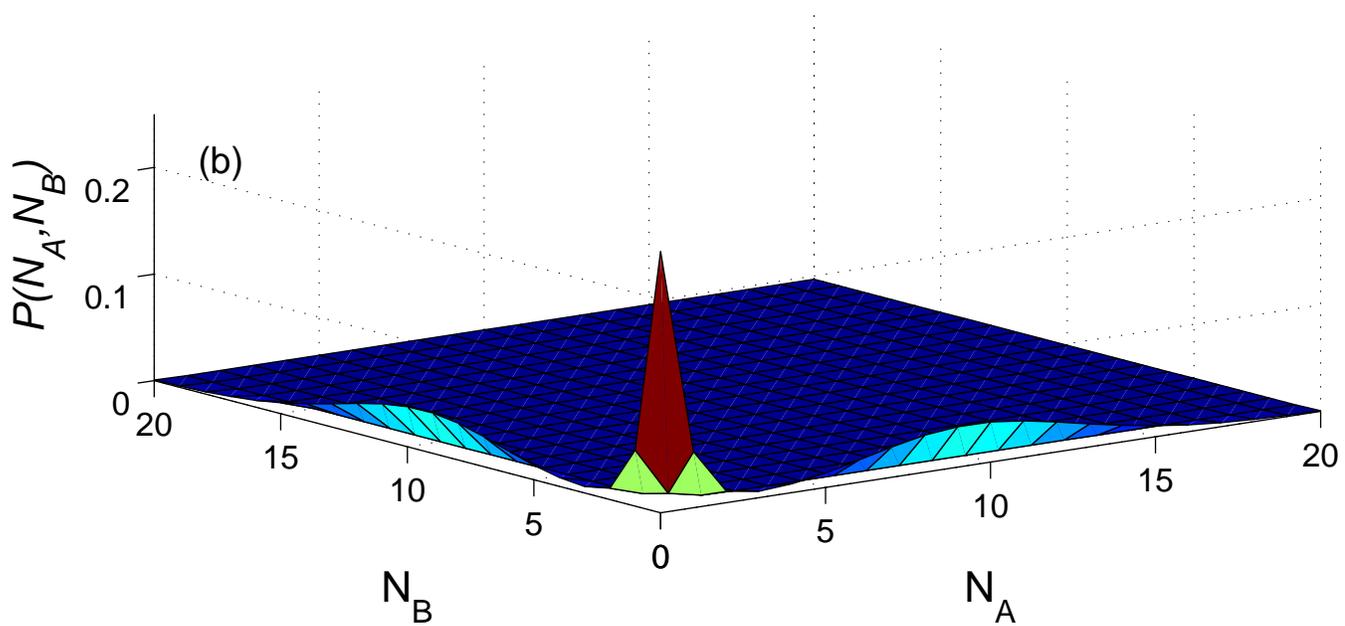
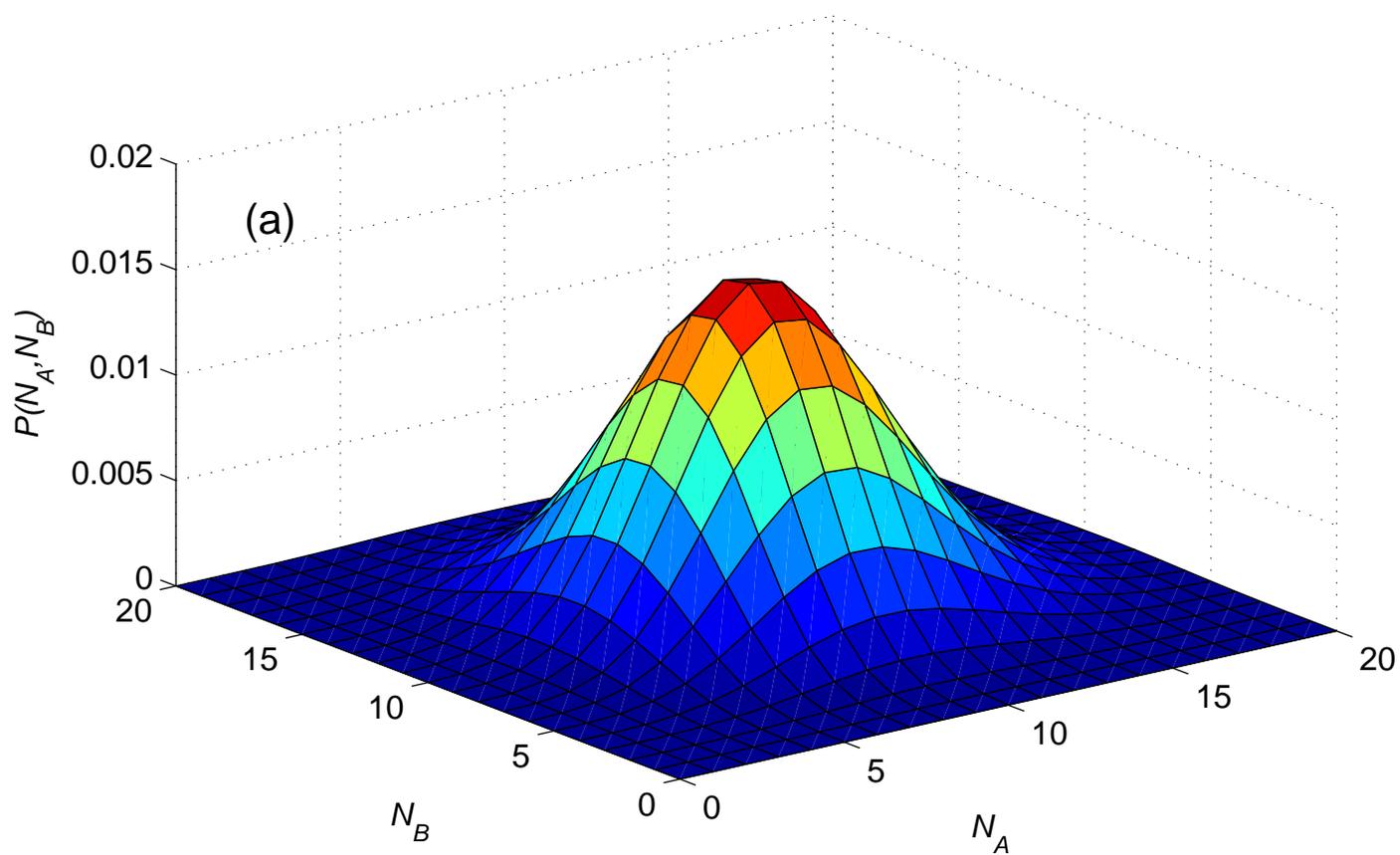
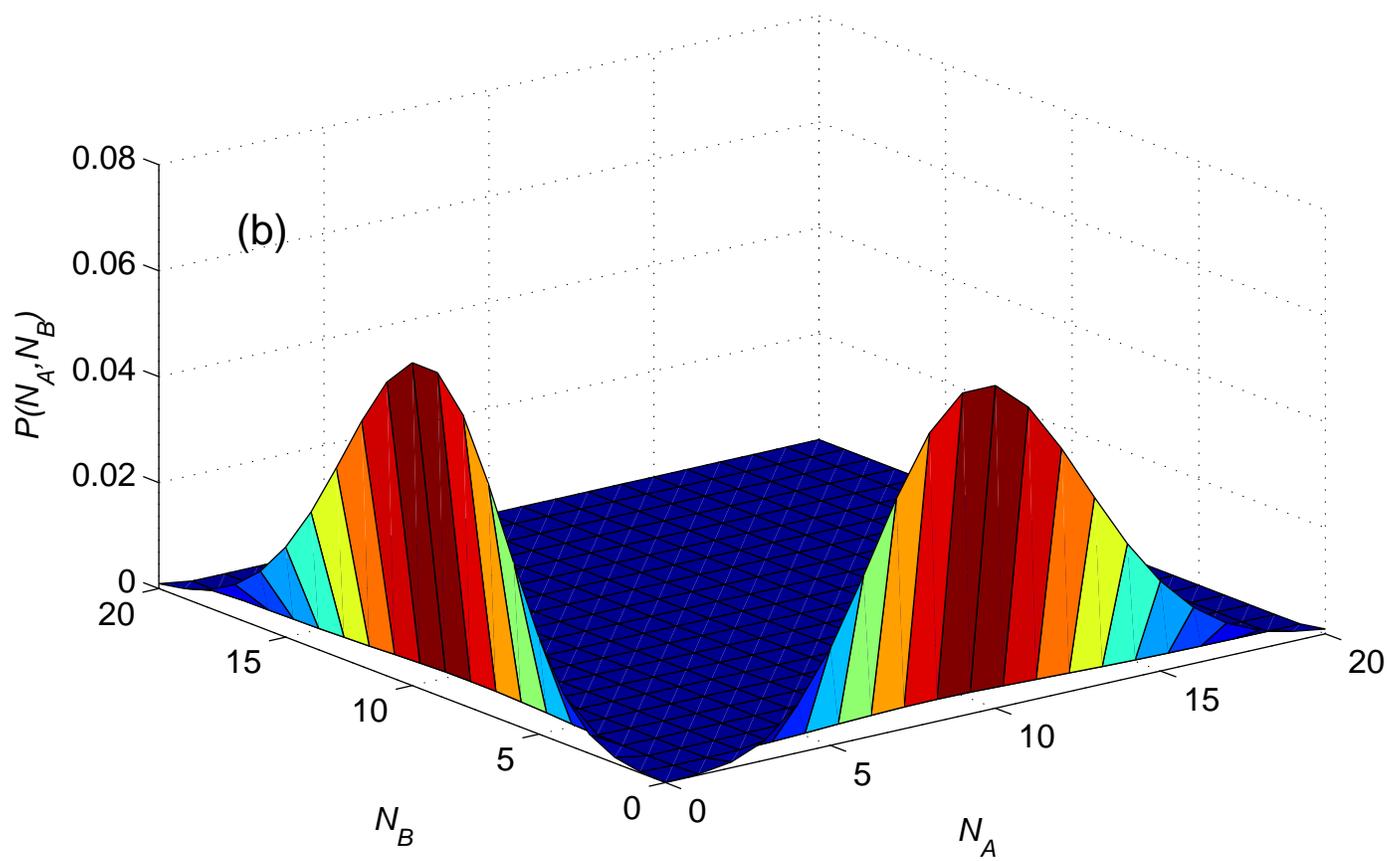
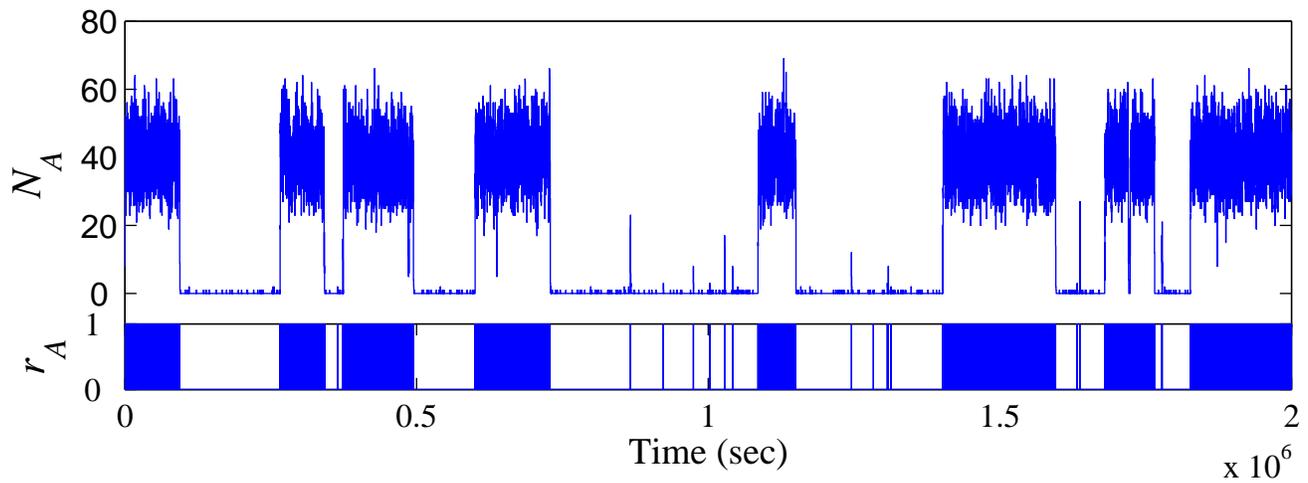
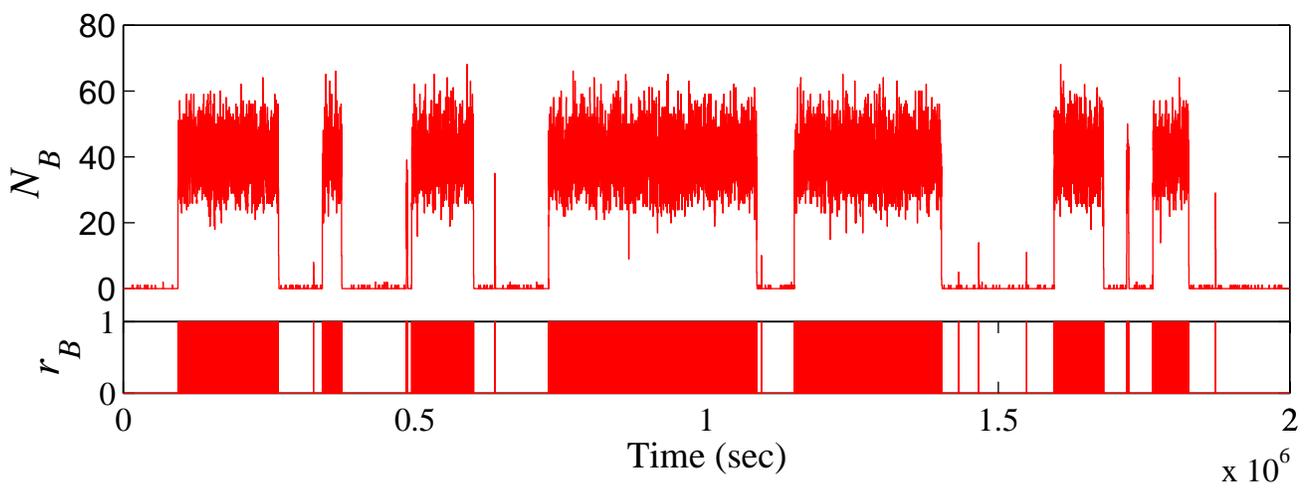
\end{document}